\documentclass[aps,pre,twocolumn,%
superscriptaddress,letterpaper,10pt]{revtex4-1}
\pdfoutput=1
\usepackage{amsmath,amsfonts,amssymb,amsbsy,amscd,amsgen}\usepackage[usenames,dvipsnames]{xcolor}
\usepackage{graphicx}

\newcommand{\Q}{\tau_z} 

\usepackage[normalem]{ulem}

\begin{document}

\title{Bypass transition and spot nucleation in boundary layers}

\author{Tobias Kreilos}
\email{tobias.kreilos@epfl.ch}
\affiliation{Emergent Complexity in Physical Systems Laboratory (ECPS), \'Ecole Polytechnique F\'ed\'erale de Lausanne, CH-1015 Lausanne, Switzerland}
\affiliation{Fachbereich Physik, Philipps-Universit\"at Marburg, 35032 Marburg, Germany}
\author{Taras Khapko}
\affiliation{Linn\'e FLOW Centre, KTH Mechanics, Royal Institute of Technology, SE-100 44 Stockholm, Sweden}
\affiliation{Swedish e-Science Research Centre (SeRC), Sweden}
\author{Philipp Schlatter}
\affiliation{Linn\'e FLOW Centre, KTH Mechanics, Royal Institute of Technology, SE-100 44 Stockholm, Sweden}
\affiliation{Swedish e-Science Research Centre (SeRC), Sweden}
\author{Yohann Duguet}
\affiliation{LIMSI-CNRS, Universit\'e Paris-Sud, Universit\'e Paris-Saclay, F-91405 Orsay, France}
\author{Dan S. Henningson}
\affiliation{Linn\'e FLOW Centre, KTH Mechanics, Royal Institute of Technology, SE-100 44 Stockholm, Sweden}
\affiliation{Swedish e-Science Research Centre (SeRC), Sweden}
\author{Bruno Eckhardt}
\affiliation{Fachbereich Physik, Philipps-Universit\"at Marburg, 35032 Marburg, Germany}
\affiliation{J.M. Burgerscentrum, Delft University of Technology, NL-2628 CD Delft, The Netherlands}

\date{\today}

\begin{abstract}
The spatio-temporal aspects of the transition to turbulence are considered in the case of a boundary layer
flow developing above a flat plate exposed to free-stream turbulence. Combining results on the receptivity
to free-stream turbulence with the nonlinear concept of a transition threshold, a physically motivated
model suggests a spatial distribution of spot nucleation events. 
To describe the evolution of turbulent spots a probabilistic cellular automaton
is introduced,
with all parameters directly fitted from numerical simulations of the boundary layer.
The nucleation rates are then combined with the cellular automaton model, yielding excellent quantitative agreement with the statistical characteristics
for different free-stream turbulence levels.
We thus show how the recent theoretical progress on transitional wall-bounded flows can be
extended to the much wider class of spatially developing boundary-layer flows.
\end{abstract}

\maketitle

\section{Introduction}

The boundary layers that form whenever a fluid flows over a solid surface determine many physical properties
such as the drag on the surface or the transfer of heat \citep{Schlichting2004}. The theory for the laminar
boundary layer was developed by Prandtl and Blasius, who described the velocity 
profile and the characteristic downstream variation of the boundary layer. The transition to a turbulent boundary
layer is accompanied by dramatic changes in its physical properties, and remains a fascinating object of study because
it often does not follow the linear instability described by Tollmien and Schlichting. Instead, finite amplitude
perturbations can trigger turbulence much more quickly in a process dubbed bypass transition, so named to indicate
that it circumvents the linear instability~\citep{schmid2001}.
The transitional region of the boundary layer is characterized by spatially and temporally fluctuating 
turbulent spots with an increasing probability to be turbulent farther downstream \citep{Emmons1951,Klebanoff1961}.

A key quantity in the characterization of the transition is
the intermittency factor $\gamma(x)$, defined as the 
probability to be turbulent at streamwise position $x$.
Most models that have been developed for $\gamma$ contain
phenomenological assumptions about the nucleation rate of spots and their further
evolution \citep{Emmons1951,Dhawan1957,Narasimha1985,Johnson1999,Vinod2004,Vinod2007}. 
An exception is the model described in \citep{Ustinov2013}, where
transient amplification of perturbations and a threshold for the transition
are used to derive a dynamical model for the spot nucleation rate and hence
$\gamma$. Other properties of the dynamics, such as the
number and width of turbulent regions, are not considered.
The model we describe here is based on our understanding of the transition
in internal flows and contains a cellular automaton representation
of the dynamics that also captures the time evolution of the spots.

The transition to turbulence in parallel flows such as plane Couette flow or pipe flow \citep{Wygnanski1973,Lundbladh1991,Avila2011} shares many 
features with bypass transition in the spatially developing boundary layer
\citep{Emmons1951,Vinod2004}.
In both sets of flows
the laminar profile is conditionally stable and finite perturbations are needed to trigger the transition.
In the case of parallel flows, 
the transition to turbulence has been linked to the appearance of \mbox{3-D} exact coherent structures via 
saddle--node bifurcations and their connections in the global state space of the system 
\citep{Faisst2003,Eckhardt2007,Eckhardt2008b}. The boundary between laminar and 
turbulent motion, defined by the singularities in lifetime measurements,
is formed by the stable manifold of the so-called edge state, which determines the threshold needed to trigger 
turbulence \cite{Skufca2006}.
As a step towards identifying this key feature in boundary layers the edge trajectory intermediate between 
laminar and turbulent dynamics has been computed in \citep{Duguet2012,Cherubini2011}. 
Compared to the parallel internal flows, the spatial development of the boundary layer changes 
the scale of the structures as one moves downstream, but it is clear that the initial condition has to 
pass a certain threshold in the inflow region in order to become turbulent. Optimal flow
structures for the transition and their subsequent temporal and spatial development have been discussed in 
\cite{Cherubini2011b,Duguet2013a,Kerswell2014}. 

In this paper we show how the concept of an edge state and its instability
can be used to derive a model for the nucleation of turbulent spots in the boundary
layer subject to free-stream turbulence (FST). This model is then combined with a
probabilistic approach to turbulent spreading to obtain a physics-based model
for the birth and evolution of localized spots.

\section{Numerical data} \label{sec:num}

The model developed in this paper is designed quantitatively from
numerical data. Simulations of the incompressible Navier--Stokes equations in a
Blasius geometry, under the influence of free-stream turbulence, have been performed using the spectral code SIMSON
\citep{simson,Schlatter2004}. These simulations have
been shown to be in very good agreement with experimental observations~\citep{Orlu2013}.

In parallel flows the flow rate and the characteristic length are usually constant. 
In the spatially developing Blasius boundary layer only the free-stream
velocity $U_\infty$ is constant while the thickness $\delta(x)$ increases in the downstream ($x$-)direction
(specifically, we define $\delta(x)$ as the displacement thickness \citep{Schlichting2004}).
Accordingly, the Reynolds number $R(x)$ varies in space, $R(x)=U_\infty \delta(x)/\nu=1.72\sqrt{U_\infty x/\nu}$ 
(with $\nu$ the kinematic viscosity). 

The computational domain
starts at a distance $x_0$ from the edge of the plate with $R(x_0) = 300$.
In units of the displacement thickness $\delta_0$ at this location, $x_0 = 101$ and the domain has dimensions $2000 \times 130 \times 500$ in the
downstream $x$, wall-normal $y$ and spanwise $z$
direction.
At the end of the domain, a fringe region is introduced in which the
perturbations are damped and returned to the Blasius profile. 
Further details of the numerical code can be found in Ref.~\citep{simson,Schlatter2004}. 
More details on the numerical parameters and the simulations are given in the appendix.

\begin{figure}
  \includegraphics[width=\linewidth]{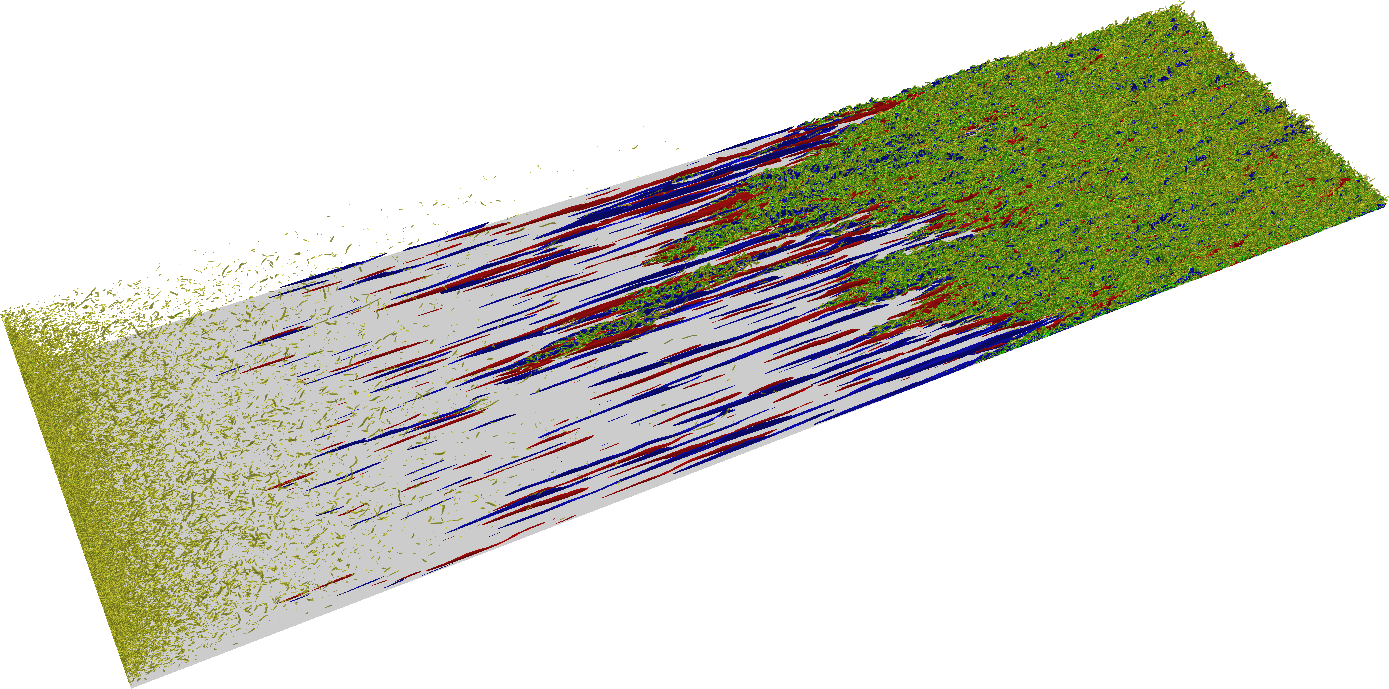}\\
  \includegraphics[width=\linewidth]{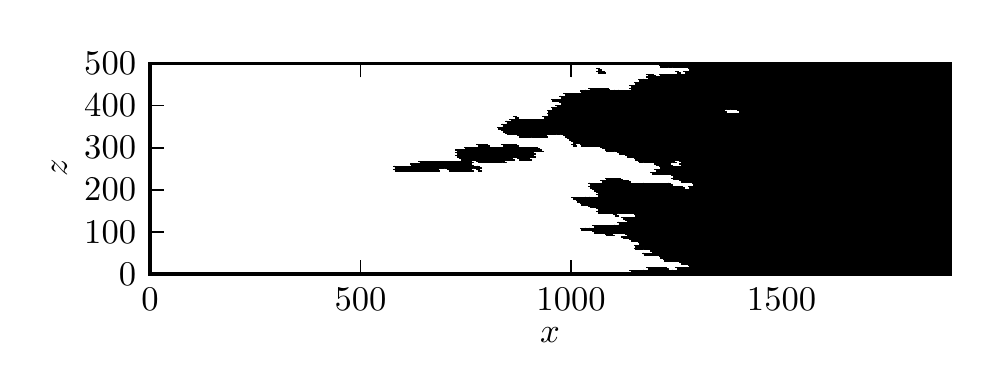}
  \caption{\label{fig:simulation}
    Two levels of representation of the turbulence transition in boundary layers. Top: a snapshot from a numerical solution
    of the full equations (movies are supplied with the supplemental material). 
    Free-stream turbulence enters from the left (vortices visualized in green by isocontours of $\lambda_2$).
    As it moves along the plate, it decays and 
    induces perturbations in the boundary layer which develop into low- and high-speed streaks (blue and red, respectively)
    that then break down and initiate turbulence (green
    regions to the right) that grow and spread to fill the boundary layer.
    Bottom: reduction of the above snapshot to a discrete laminar/turbulent representation according to the local spanwise wall shear.
    Turbulent regions are black, laminar ones white.
  }
\end{figure}

A snapshot from a numerical simulation in Fig.~\ref{fig:simulation} (top) shows several stages
of the flow development from the initial perturbations upstream through the emergence of
streaks and their breakdown into isolated turbulent spots that grow to cover the entire width of the domain further downstream.
The intermittency factor $\gamma$ depends on the turbulence intensity, characterized by the parameter
$Tu = \sqrt{(u_{rms}^2 + v_{rms}^2 + w_{rms}^2)/3}$ in units of $U_\infty$.
We focus on the range of $Tu$ between $3\%$ and $4\%$, well inside
the region $Tu \gtrsim 2 \%$ where bypass transition typically occurs.

The original simulation data is transferred to a coarser Cartesian grid defining the individual cells for the model.
We furthermore neglect variations in the wall-normal direction and reduce the boundary layer to two dimensions, an approach that is justified by many
experimental and numerical studies.
 As a local indicator for turbulence the local spanwise shear
stress at the wall $\Q = \partial w / \partial y|_{y=0}$ is used. Before transition to turbulence, the flow consists mainly of streamwise oriented streaks, which have high energy in the downstream velocity fluctuations but only very little in the spanwise ones.
 After breakdown of the streaks, the flow exhibits strong vortical motion. Strong streamwise vortices lead to a higher spanwise wall-shear stress, so
that $\Q$ is high if the flow is turbulent. Furthermore, $\Q$ is a wall-based quantity, showing no ambiguity in the position where it is measured
and monitored from the numerical simulations.

\begin{figure}
    \includegraphics[width=\linewidth]{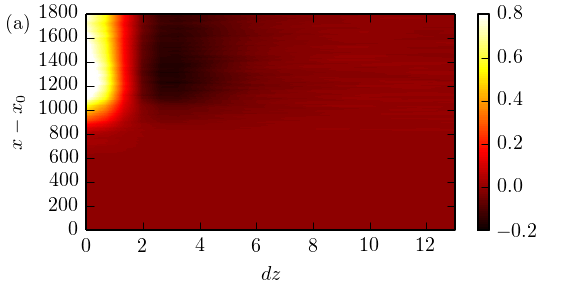}
    \includegraphics[width=\linewidth]{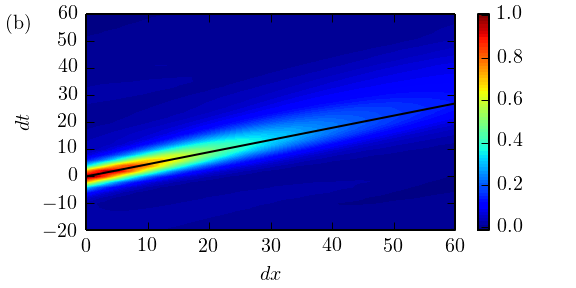}
    \caption{\label{fig:autocorr}
      (a) Autocorrelation in the spanwise direction for every downstream location $x$. Before the transition region
      $\Q$ vanishes, afterwards the autocorrelation function is almost independent of $x$. There is a strong
      positive correlation for $z \lesssim 2$ and a somewhat weaker, but clear, anticorrelation for $2 \lesssim z \lesssim 5$,
      corresponding to a vortex and the counter-rotating neighbor, respectively.
      (b) The autocorrelation function of $\Q$ in time and downstream direction shows a strong positive correlation in the
direction
    $dx/dt = 0.45$, indicated by the black line and corresponding to the average advection speed in the boundary layer.
    }
  \end{figure}

  The grid spacing of the numerical simulations is $D_x = 1.95$ and $D_z = 0.65$
  in units of $\delta_0$.
  For the probabilistic model, we have to determine the size of independent cells and a suitable time step. To get an estimate of an appropriate
discretization, we look at the autocorrelation function
  of $\Q$. Since we expect the structures to be advected quickly in the downstream direction, but only slowly in the spanwise one,
  we calculate the purely spatial autocorrelation in the spanwise direction (Fig.~\ref{fig:autocorr} a)
  and the space-time autocorrelation in the downstream direction (Fig.~\ref{fig:autocorr} b).

  The autocorrelation in the spanwise direction is computed independently for all downstream positions,
  $C_z(x,dz) = \left\langle\int \Q(x,z,t)\Q(x,z+dz,t)\mathrm dz\right\rangle_t$, with $\langle\rangle_t$ indicating temporal
averaging. Figure~\ref{fig:autocorr}(a) shows that it is extremely small before transition to turbulence occurs. Afterwards, it is almost
independent of
$x$, indicating that the size of the structures does not depend on the downstream location.
  There is a strong positive correlation for $z \lesssim 1.5$, corresponding to the width of a single vortex,
  and a somewhat weaker but still clear negative correlation for $2 \lesssim z \lesssim 5$, corresponding to
  the counter-rotating vortex. As we want our cell size to average over one vortex pair, which ranges from
  $-2$ to $5$, a good estimate of $dz$ is hence given by $dz \simeq 6$-$7$ and we choose $dz = 10 D_z = 6.5$ so that it is an integer multiple of the
grid spacing in the numerical simulations.

  Looking at the space-time autocorrelation 
  \[C_{xt}(dx,dt) = \left\langle\int \Q(x,z,t)\Q(x+dx,z,t+dt)\mathrm dx \mathrm dt\right\rangle_z\]
  in Fig.~\ref{fig:autocorr}(b),
  we see a very strong positive finger
  pointing into the plane, corresponding to the speed at which the structures are advected. The finger is
  rather thin, indicating that the advection speed is constant everywhere for all structures. The finger 
  has a slope of $dx/dt = 0.45$, which is depicted by the black line and we
  naturally choose this measure to define $dt$ once $dx$ is chosen.
  The autocorrelation function, however, does not give a clear estimate for $dx$ and we deliberately choose
  $dx = 5D_x = 9.5$ as a compromise between averaging over enough gridpoints and keeping the time step low (which
  means more statistics from a simulated trajectory). The time step that follows is $dt = dx/0.45 = 22$.
  We have tried different values for $dx$ during the fitting procedure outlined below and verified a posteriori that the exact choice of $dx$ does not
influence our results, \emph{e.g.}\ for the intermittency factor,
as long as $dx$ is not too large. Note, however, that the turbulence spreading parameters
discussed in the next section do depend on $dx$ and have to be adjusted accordingly.

  The 3D box size of the simulations $L_x \times L_y \times L_z = 2000 \times 130 \times 500$ translates to a 2D cell grid of size $N_x\times N_z =
204\times76$ for
  the model. The data is reduced to a coarser grid using local spatial averaging.

\begin{figure}
    \includegraphics[width=\linewidth]{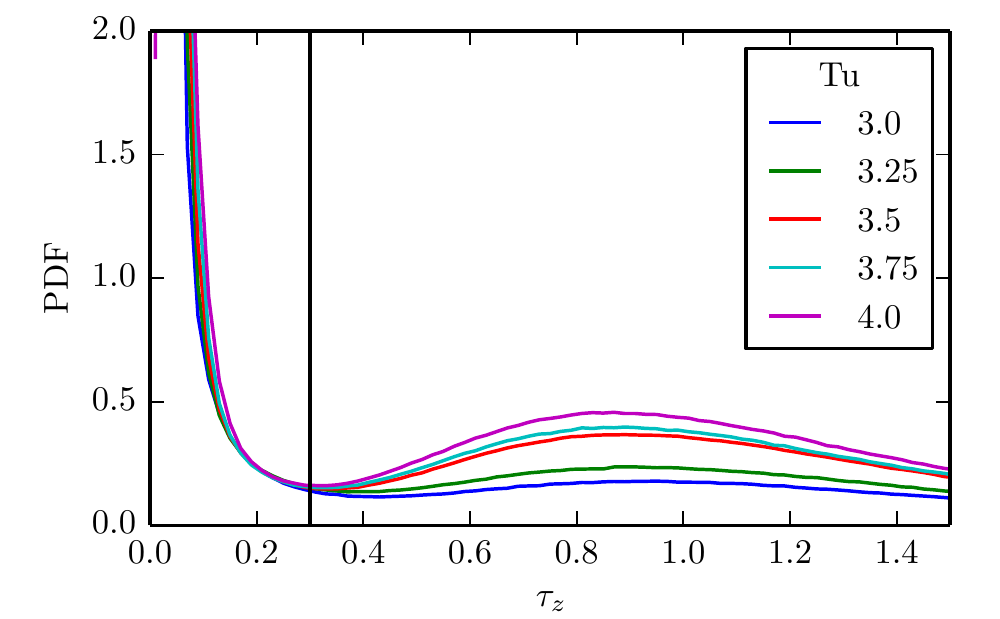}
    \caption{\label{fig:threshold}
      The probability density function of $\Q$ for different FST intensities is used
      to determine a threshold defining laminar and turbulent. The chosen threshold is located close to the minimum between the two peaks and is indicated by the black vertical line at $0.3$.
    }
  \end{figure}

In order to distinguish between laminar and turbulent cells we choose a threshold for $\Q$ and define
  everything below the threshold as laminar and everything above it as turbulent.
  The threshold is estimated from the probability density function of $\Q$, shown in Fig.~\ref{fig:threshold}
  for all five turbulence intensity levels.
  The PDF is high near~$0$, drops to a minimum and then shows a peak, whose height increases with free-stream
  turbulence intensity as larger parts of the box are turbulent.
  Associating the high values near $0$ with patches of purely laminar flow and the second peak at higher values of the spanwise wall-shear stress with
  turbulent patches, we set the threshold in the gap separating the two at $\Q = 0.3$. 
  
  After applying the threshold a few undesired effects remain: we sometimes find a single laminar cell 
  in a turbulent region or a flickering of isolated turbulent cells in a laminar region that appear for a single
  time step only. To prevent those spurious events from contaminating our statistics, we
  apply a Gaussian filter with kernel size $0.5$ cells in both spatial directions before applying the threshold.

  The final result of our data processing procedure is shown in Fig.~\ref{fig:simulation} (bottom), where the 
  2D binary representation of the above snapshot is shown. The figure suggests
  that our criterion captures the location of turbulent patches (green in the upper 
  snapshot and black in the lower one) very well.

\section{Modelling spot evolution}

\begin{figure}
  \includegraphics[width=.9\linewidth]{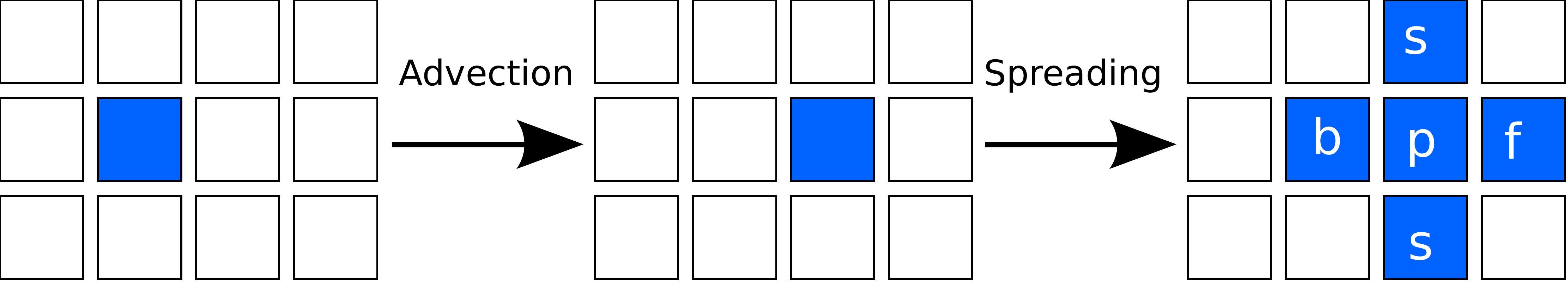}\\
  \caption{\label{fig:pcamodel}
    Reduction of the spatial and temporal dynamics on a discrete lattice of cells that are either laminar (white) or turbulent
    (blue). The temporal evolution of a turbulent region consists of advection by one cell to the right with persist probability $p_p$, and
 spreading to
    neighboring cells with probabilities $p_b$, $p_s$ and $p_f$. 
  }
\end{figure}

The simulations, both in the full representation as well as in their reduced binary description, show the nucleation
of turbulent spots at spatially and temporally varying positions at the upstream side, and their advection
and growth in the downstream direction. 
We here focus on the evolution of turbulent spots, which we describe using probabilistic
cellular automata (PCA) \citep{Chate1988,Daviaud1990,Barkley2011,Allhoff2012}.

With the discretization of space and time discussed in section~\ref{sec:num}, we now look for a
discrete dynamics that updates the state of each cell.
Each temporal update in the probabalistic cellular automaton follows two steps.
The first deterministic step models the advection, translating all cells by one unit in the downstream direction.
  In a second step, the cell can spread or decay. The probabilities are $p_f$ to spread forward, $p_p$ to persist, $p_s$ to
  spread right or left and $p_b$ to spread backwards, as shown in Fig.~\ref{fig:pcamodel}.
  
  The numerical values of the four probabilities are directly extracted from the numerical
  data in the following way:
  the probability that a cell $C$ is laminar after one time step, $p_{l}=p( C(x+1, z, t+1) \equiv 0)$
  is given by the product of the probabilities that the surrounding cells do not spread turbulence
  in this cell and reads:
  \begin{align*}p_l=(1-p_p) C(x,z,t) \cdot (1-p_b) C(x+1,z,t) \\ \cdot (1-p_f) C(x-1,z,t) \cdot (1-p_s) C(x,z-1,t) \\ \cdot (1-p_s) C(x, z+1, t)\end{align*}
  Measuring $p_l$ for all possible configurations of surrounding cells in the numerical data,
  we obtain a system of equations from which the probabilities can be calculated using a least-squares algorithm.
  
\begin{figure}
  \includegraphics[width=\linewidth]{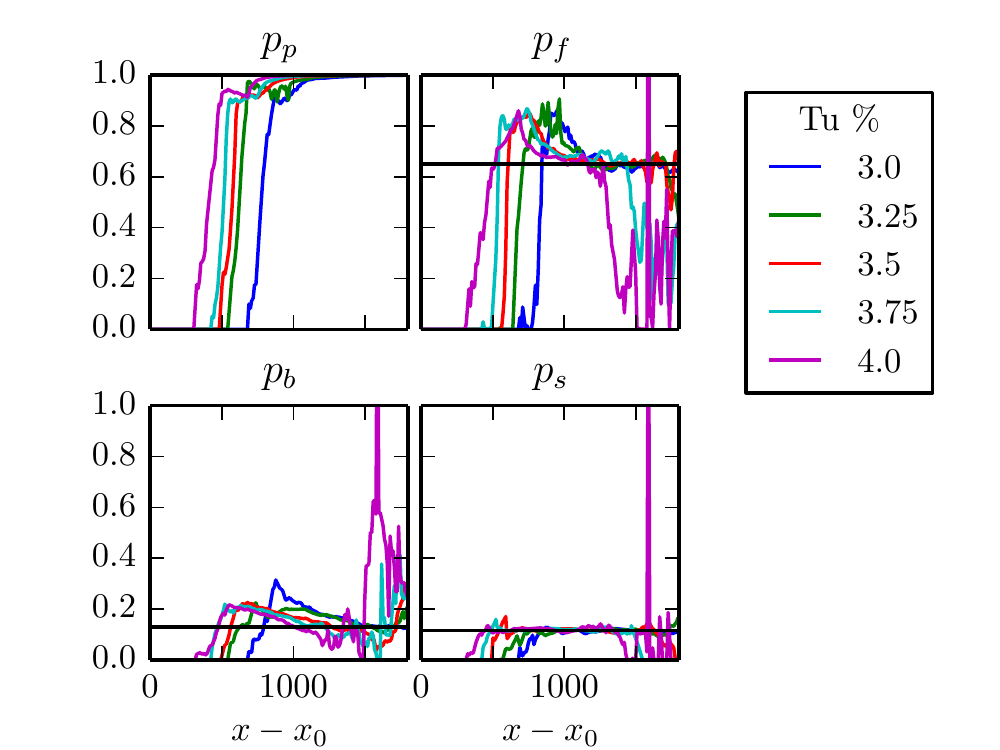}
  \caption{\label{fig:parfits}
    Probabilities to persist ($p_p$), spread forward ($p_f$), sideways ($p_s$) or backwards ($p_b$) estimated
    from the
    numerical data for different FST intensities. While there is some variation in the earliest phase of transition,
    where very
    few events happen and the statistics are poor, there is afterwards almost no variation with downstream position.
    Furthermore,
    there is no dependence of the probabilities on the level of FST disturbances -- once a turbulent spot is created,
    its time
    evolution is intrinsic and independent of position and what happens in the free stream.
    The black lines indicate the constant values of the probabilities that are chosen for the PCA, $p_p$ is equal to one.
  }
\end{figure} 

Figure~\ref{fig:parfits} shows the resulting probabilities for all FST intensities.
The probabilities show strong similarities for all Tu-levels, with a sharp increase near the onset of transition and a quick 
settling to an almost constant value afterwards, with $p_f$ and $p_b$ showing a slight overshoot near the onset.
Disregarding the laminar region before any turbulence is encountered, and both onset and late stages of transition, where
almost no events are detected during the simulations and the statistics is extremely poor, all probabilities appear to be almost
independent of both $R$ and $Tu$.
We therefore choose constant probabilities for the PCA, the values are indicated by the black lines in Fig.~\ref{fig:parfits}.
Note that $p_p=1$, so that there is no significant spontaneous relaminarization inside a turbulent cell.
It is worth noting that the development of turbulent spots in the transitional boundary layer
can hence be described as an activated process, with the properties describing the spot
evolution being independent of $R$ and $Tu$.

The probabilistic model is simulated on the cells corresponding to the coarsened grid of the numerical simulation, with $N_x\times N_z = 204\times76$
cells, spanwise periodicity and an unperturbed inflow.

\section{Modelling spot nucleation}

To obtain a complete description of the evolution of spots in the boundary layer, we need to supplement the spreading process with a position-dependent 
rate for the nucleation of new turbulent spots, $p_c(x)$, which enters the 
cellular automaton as the probability per unit time to have a nucleation event in a cell
at position $x$.

\begin{figure}
  \includegraphics[width=\linewidth]{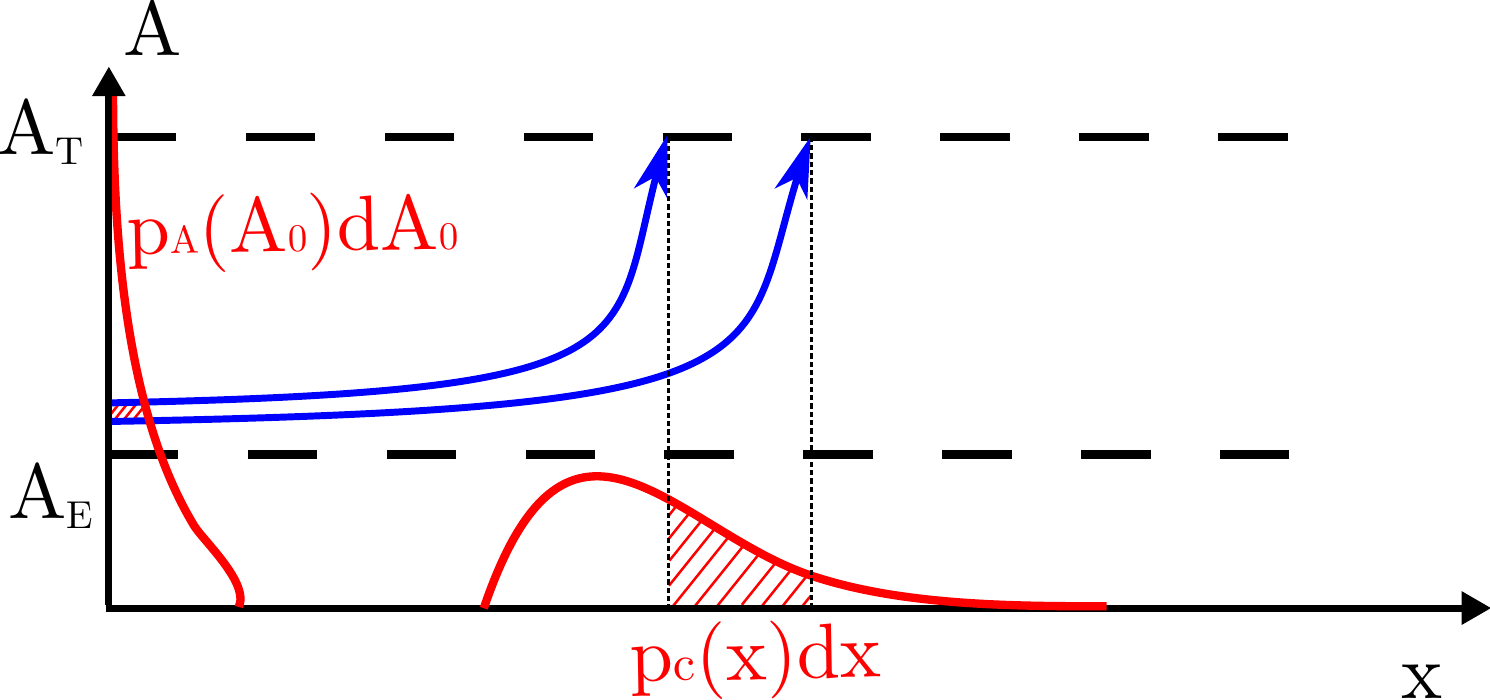}
  \caption{\label{fig:model}
   The basic processes underlying the transition in boundary layers: perturbations of some initial amplitude $A_0$ enter the boundary layer. 
    If the initial perturbations are above the threshold $A_E$ associated with the stable manifold of the edge state, they exponentially grow away from the edge (blue arrows), until they
    reach a threshold $A_T$ where they trigger the creation of a turbulent spot. Since they are advected downstream while growing, we can
    assign a transition location to each initial amplitude. A distribution of initial amplitudes $p_A({A_0})$ hence translates into a distribution
    of spot nucleations $p_c(x)$ (red curves), allowing us to overcome the hypothesis of concentrated breakdown \citep{Dhawan1957}.
  }
\end{figure}

The physical process underlying the nucleation of turbulent spots is the response of the
boundary layer to perturbations from the free-stream turbulence. Perturbations from the FST 
develop streaks that grow in intensity until they break down via secondary instabilities and initiate turbulence \citep{Andersson2001,Matsubara2001,Brandt2004,Fransson2005,Schlatter2008,Shahinfar2011}.
As in many experiments, in the numerical simulations that form the basis
of our study the flow is continuously perturbed upstream and then advected downstream. Accordingly,
the downstream development of the flow is a consequence of the time evolution of \emph{initial} conditions prepared 
upstream. If the amplitude $A$ of an initial condition is below the threshold defined by the stable manifold of the edge state, the perturbation can
be expected to decay. On the other hand, if it is sufficiently strong, it will grow exponentially fast and 
eventually trigger turbulence (Fig.~\ref{fig:model}). This simple nucleation model neglects spatial interactions and assumes constant energy
level of the edge, 
which is sufficient for quantitatively accurate predictions of the location of spots and their statistical properties, as will be shown now.

A prediction for the nucleation probabilities is obtained from the following hypotheses:
(i) time and downstream location can be used interchangeably following a standard Taylor's hypothesis;
(ii) the amplitude of the initial condition $A_0$ has to exceed a threshold $A_E$ (related to the edge) in order to lead to the nucleation of any turbulence
at all; 
(iii) since the edge is linearly unstable, the difference $A(x)-A_E$ will start to grow exponentially if the perturbation
is larger then $A_E$:
\begin{equation}
A(x) = A_E + (A_0-A_E) \exp (\lambda x)
\label{Eq1}
\end{equation}
with a Lyapunov exponent $\lambda$;
(iv) turbulence is triggered once the perturbation has reached a certain amplitude $A_T$.
Solving Eq.~(\ref{Eq1}) for $A_0$ and substituting
$A(x)=A_T$, we find
\begin{equation}
  A_0(x) = A_E + (A_T - A_E)\exp(-\lambda x)
\end{equation}
and can then translate the distribution of 
initial amplitudes $p_A(A_0)$ into the distribution of  nucleation events $p_c(x)$, viz.
\begin{equation}
p_c(x)dx = p_A(A_0(x)) \left| \frac{d A_0}{dx} \right| dx\,,
\end{equation}
which leads to:
\begin{eqnarray}
p_c(x) = p_A(A_E+(A_T-A_E) \exp(-\lambda x)) \times \\\nonumber \lambda (A_T-A_E) \exp(-\lambda x) \,.
\end{eqnarray}

The initial fluctuations are assumed to be Gaussian, so that
\begin{equation}
p_A(A_0) = \exp(-A_0^2/\sigma^2)/(\sqrt{\pi} \sigma) \,,
\end{equation}
where the standard deviation $\sigma$ increases with the turbulence level $Tu$ \citep{Andersson2001}.
Then
\begin{eqnarray}
p_c(x) = \exp\left(-\left(\frac{A_E+(A_T-A_E) \mathrm e^{-\lambda x}}{\sigma}\right)^2-\lambda x\right) \times \\\nonumber \lambda
(A_T-A_E)/(\sqrt{\pi} \sigma) \,.
\end{eqnarray}

This expression has several parameters:
(i) the standard deviation $\sigma$,
(ii) the Lyapunov exponent $\lambda$, 
(iii) the ratio between the threshold and the edge, $r=A_T/A_E$.
The parameters are fixed by fitting $\gamma$ determined from the time evolution of the 
cellular automaton using the modeled nucleation rate to $\gamma$ determined in the numerical
simulations. The comparison shows that a good fit can be obtained with a constant $r\gg1$, which justifies neglecting the fluctuations of
the edge amplitude. The relation between $\sigma$ and $Tu$ appears to be linear. The fit also
reveals a linear increase of the growth rate $\lambda$ with $Tu$.
The latter is interpreted by the observation that higher $Tu$ leads to stronger streamwise vortices
in the boundary layer, which give rise to a faster growth of the streaks \citep{Fransson2005}.
For the final fit, we imposed functional relations
and determined the parameter values indicated in Table~\ref{tab:parameters}.
The finally obtained linear relations are $\sigma = 0.226\; Tu_\% - 0.08$ and $\lambda =  (5\; Tu_\% - 8) \cdot 10^{-3}$.

\begin{table}[h]
\centering
\caption{\label{tab:parameters} Parameters of the nucleation model for different turbulent intensities $Tu$.}
\begin{tabular*}{\linewidth}{@{\extracolsep{\fill} } c|c|c|c|c|c}
\hline
\hline
Parameter & $3.0\%$ & $3.25\%$ & $3.5\%$ & $3.75\%$ & $4.0\%$ \\
\hline
$r$ & 145 & 145 & 145 & 145 & 145 \\
$\sigma$ & 0.60 & 0.66 & 0.71 & 0.77 & 0.82 \\
$\lambda \times 10^3$ & 6.79 & 8.05 & 9.32 & 10.6 & 11.8 \\
\hline
\hline
\end{tabular*}
\end{table}

\begin{figure}
  \centering
  \includegraphics[width=\linewidth]{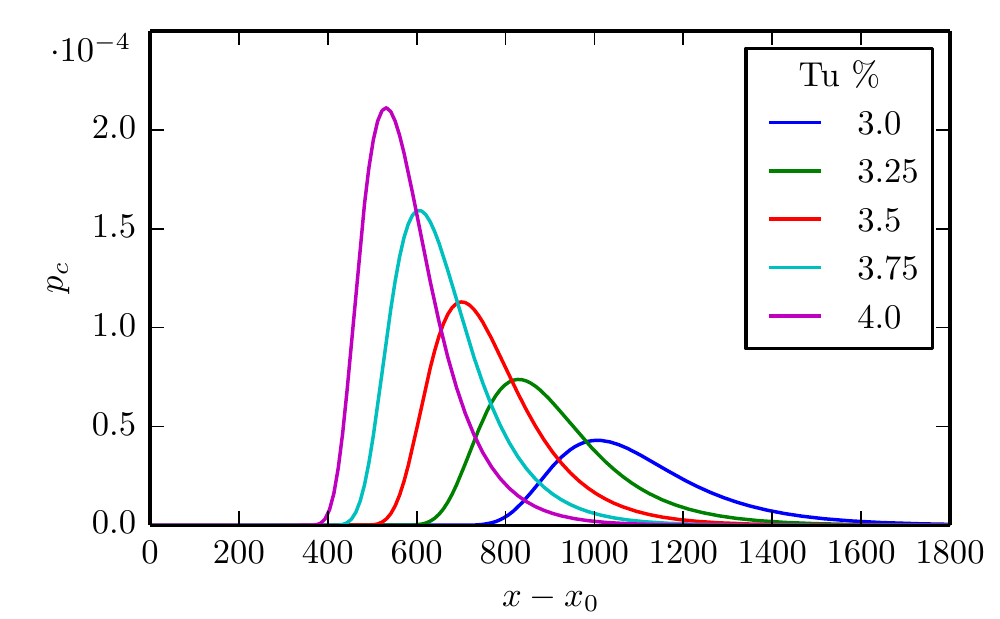}
  \caption{\label{fig:probc}
    Nucleation rate $p_c(x)$ for different values of the turbulence level $Tu$. Note the upstream motion of the maximum and the narrowing
    of the distribution with increasing $Tu$. The position $x$ is measured in units of the displacement 
    thickness $\delta_0$ at the point of entry; the offset $x_0$ marks the distance from the edge of the plate to 
    the upstream end of the numerical domain.}
\end{figure}

The obtained probability distributions $p_c$ are shown in Fig.~\ref{fig:probc} for different values of $Tu$. One notes that
they shift upstream and become narrower with increasing $Tu$. The overall shape is compatible with the
data of \citet{Nolan2013}. The rapid increase at the upstream end is a consequence
of the exponential amplification and the tail on the downstream side comes from the initial conditions that are very close
to the edge and that need more time to reach the turbulence level $A_T$.

\section{Results and discussion}

We have developed a probabilistic cellular automaton model for the evolution of turbulent
spots and a physics-inspired model for the nucleation of spots. Combining the two
the full dynamics of the boundary layer can be simulated at very low computational cost.

\begin{figure}
  \includegraphics[width=\linewidth]{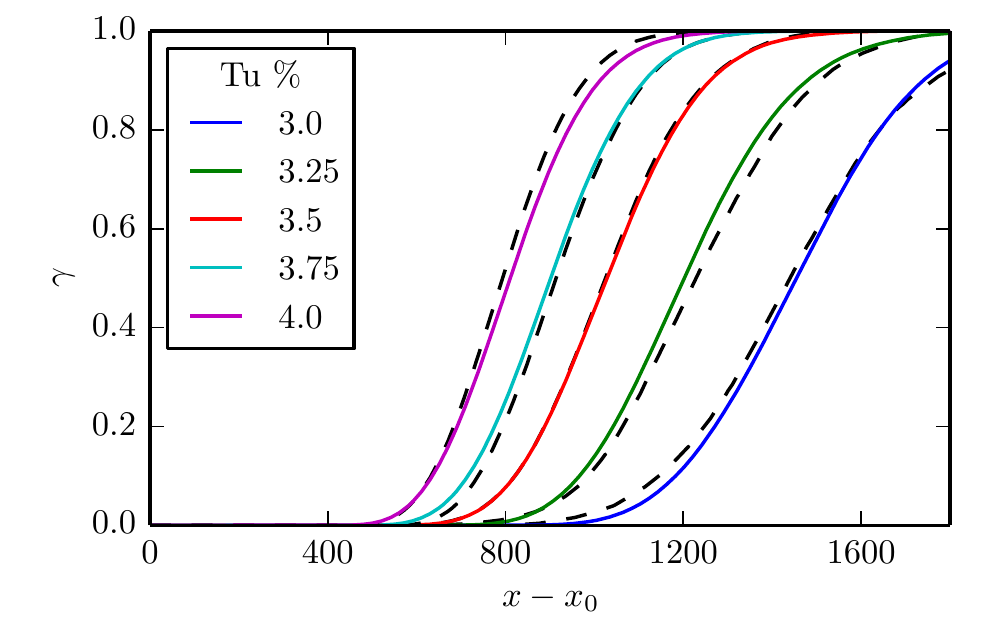}
  \caption{\label{fig:gammare}
  Comparison between the numerical simulations of the flow and the results from the cellular automaton model with
  intermittency curves $\gamma(x,Tu)$ for different turbulence levels. Black (dashed): simulation data, colours: automaton model.
}
\end{figure}

\begin{figure}
    \includegraphics[width=\linewidth]{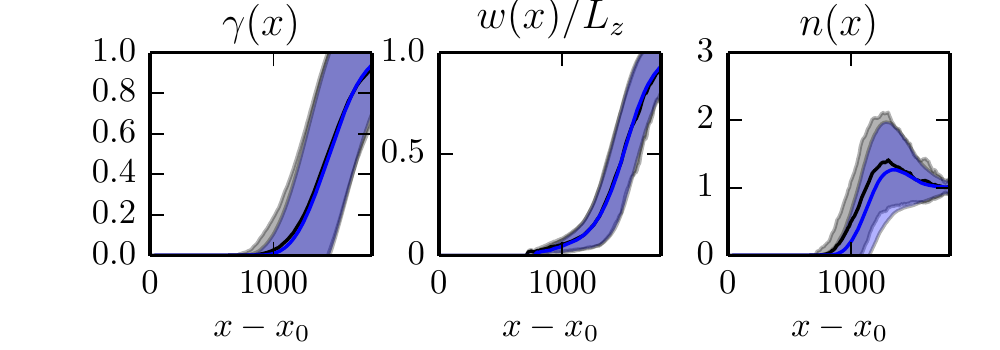}
    \includegraphics[width=\linewidth]{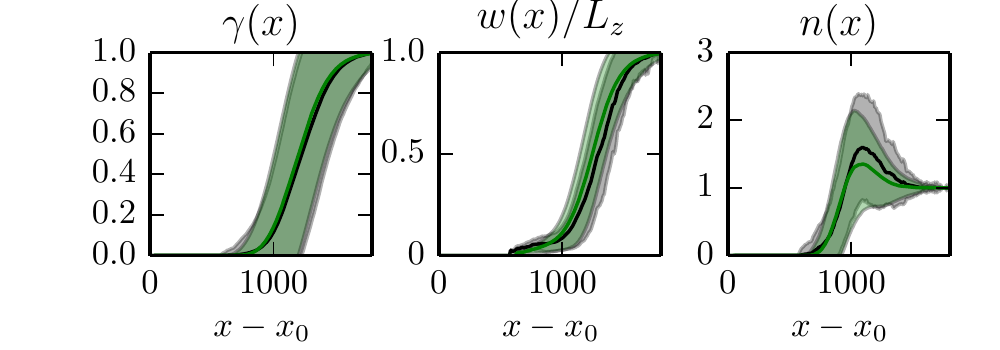}
    \includegraphics[width=\linewidth]{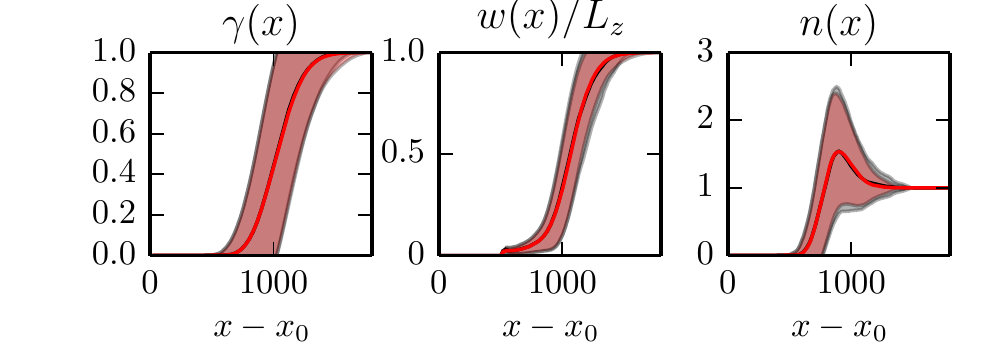}
    \includegraphics[width=\linewidth]{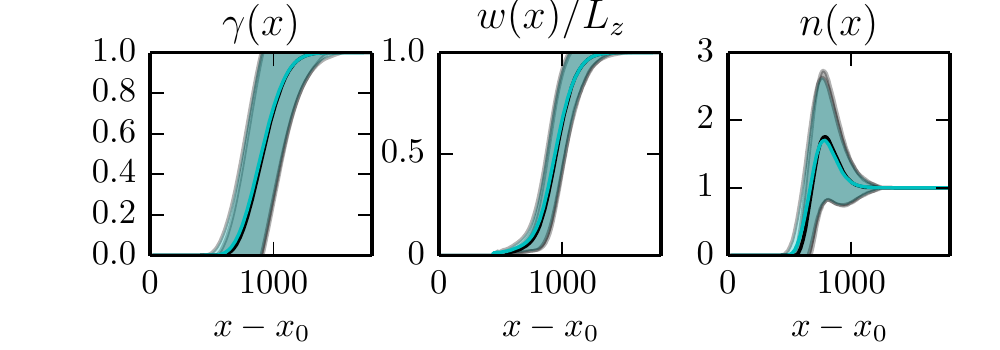}
    \includegraphics[width=\linewidth]{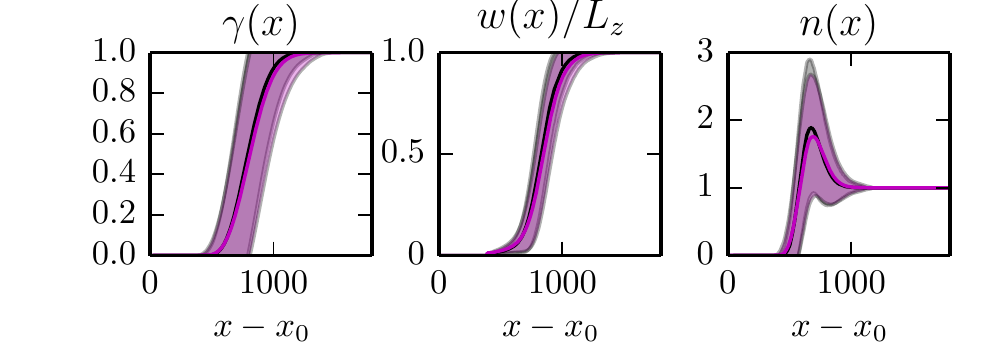}

    \caption{\label{fig:spotstats}
      Detailed comparison between the numerical data and the PCA for the five values of Tu: $3.0\%$, $3.25\%$, $3.5\%$, $3.75\%$ and $4.0\%$ (from top to bottom).
      In addition to the intermittency factor $\gamma(x)$ (left), the average width of independent spots $w(x)$ (middle) and the number of independent spots $n(x)$ (right) is shown.
      For each quantity, the value from the numerical simulations of the flow is shown in black with the gray shaded area indicating $\pm$ one 
      standard deviation and the value obtained from the cellular automaton is plotted in color.
      In most cases, the agreement is so good that no difference between the two curves is visible.
    }
  \end{figure}

As Fig.~\ref{fig:gammare}
shows, the cellular automaton model with the above nucleation rates reproduces the observed intermittency factor $\gamma$ very well.
Other quantities, such as the fluctuations around the mean (Fig.~\ref{fig:spotstats} left column), the width of individual spots (middle column)
or the number of spots (right columns) are also in very convincing agreement. We also point
to a movie (available online, see the supplementary material), comparing the numerical simulations with
our model, that shows very good visual agreement.

The results presented here show how the receptivity of the boundary layer can be combined with the nonlinear concept of a threshold curve
to explain the spot nucleation mechanism. When the nucleation model is introduced into the constructed simple cellular automaton the simulation data is fully reproduced. Note that the concentrated breakdown hypothesis that assumes a fixed location for nucleation \citep{Dhawan1957,Vinod2004} does not reproduce the data as accurately.
It is remarkable that our automaton involves only four spatially constant probabilities independent of the turbulence level.
The results are an example of how the understanding that has been obtained for parallel, internal flows can be extended to the much wider class of spatially developing boundary layers.

\subsection*{Acknowledgements}

We thank Peter Schmid for helpful comments on an earlier version of the paper and the Institute of Pure and Applied Mathematics (IPAM) at UCLA for enabling participation in the ``Mathematics in Turbulence'' program 2014.
We acknowledge the financial support from the Alexander von Humboldt Foundation. Computer time was provided by the Swedish National Infrastructure for Computing (SNIC).

\section*{Appendix: numerical simulations}

The time evolution of the boundary-layer flow is simulated using a fully spectral code \cite{simson,Schlatter2004}, which solves the incompressible Navier--Stokes equations
in an open boundary-layer geometry. For the spatial discretization of the flow field a Fourier basis is used in the streamwise $x$ and spanwise $z$ directions and a Chebyshev expansion in the
wall-normal $y$ one. Second-order Crank--Nicolson and third-order Runge--Kutta methods are used for time advancement of linear and nonlinear terms, respectively.

The no-slip (homogeneous Dirichlet) boundary conditions are imposed at the wall, whereas the free-stream is represented using Neumann
boundary
conditions. As a consequence of Fourier discretization periodic boundary conditions are imposed in the streamwise and spanwise directions. Thus in
order to simulate the spatially growing boundary layer a fringe region is included at the end of the numerical domain. In the fringe region a volume
forcing is added, damping all fluctuations and returning the flow to the required inflow state.

The entrance of the reference numerical domain is at a distance $x_0$ from the leading edge of the plate and corresponds to $R(x_0) = 300$. We measure all quantities in units of
$U_\infty$ and $\delta_0^*$ at this location. In these units $x_0 \approx 101$, and the Reynolds number, assuming laminar flow, is related to the distance from the leading edge $x$ by $R \approx 29.8 \sqrt{x}$.
We perform simulations in a box of size $L_x \times L_y \times L_z = 2000 \times 130 \times 500$ with a
resolution of $N_x \times N_y \times N_z = 1024 \times 201 \times 768$. Since our approach is based on long-time statistics, the smallest scales of turbulence are modeled by a subgrid-scale model, which reduces the computational cost.
The subgrid scales are modeled with a wall-resolved LES model of relaxation type (ADM-RT).

The free-stream turbulence at the inlet is formed by a superposition of the continuous spectrum of the Orr--Sommerfeld and Squire operators \cite{Brandt2004}.
The modes are chosen in the specific way in order to ensure isotropy of the resulting turbulence. An energy spectrum characteristic of isotropic
homogeneous turbulence is obtained by rescaling the coefficients of the superposition. The integral length scale, which corresponds to the peak
in the energy spectrum, is set to $L_I=10$. This value is somewhat higher than the ones used in Ref.~\cite{Brandt2004} and motivates the use of a higher
numerical domain in our study.

Neglecting initial transients the required simulation data is sampled over $10 000$ advective time units for $Tu=3.0\%, 3.25\%, 3.75\%$ and $4.0\%$ and for $20 000$ time units for $Tu=3.5\%$.

%

\end{document}